\documentclass[12pt,preprint]{aastex}

\shorttitle{Point Source Extraction with MOPEX.}
\shortauthors{Makovoz, D. \& Marleau, F.R.}

\begin{document}


\title{Point Source Extraction with MOPEX }

\author{David Makovoz \& Francine R. Marleau}
\affil{Spitzer Science Center, California Institute of Technology,
       Pasadena, CA 91125}
\email{davidm@ipac.caltech.edu}



\begin{abstract}          
MOPEX (MOsaicking and Point source EXtraction) is a package developed
at the {\it Spitzer} Science Center for astronomical image processing.  We
report on the point source extraction capabilities of MOPEX.  Point
source extraction is implemented as a two step process: point source
detection and profile fitting.  Non-linear matched filtering of input
images can be performed optionally to increase the signal-to-noise
ratio and improve detection of faint point sources.  Point Response
Function (PRF) fitting of point sources produces the final point
source list which includes the fluxes and improved positions of the
point sources, along with other parameters characterizing the fit.
Passive and active deblending allows for successful fitting of
confused point sources.  Aperture photometry can also be computed
for every extracted point source for an unlimited number 
of aperture sizes. PRF is estimated directly from the input
images.  Implementation of efficient methods of background and noise estimation, 
and modified Simplex algorithm contribute to the computational
efficiency of MOPEX. The package is implemented as a loosely connected set of perl
scripts, where each script runs a number of modules written in C/C++.
Input parameter setting is done through namelists, ASCII configuration
files. We present applications of point source extraction to the
mosaic images taken at 24 and 70 $\mu$m with the Multiband Imaging
Photometer (MIPS) as part of the {\it Spitzer} extragalactic First Look Survey
and to a Digital Sky Survey image.
Completeness and reliability of point source extraction is computed
using simulated data.
  
\end{abstract}


\section{Introduction}

Detection of point sources and estimation of their coordinates, fluxes and other
pertinent parameters from celestial images is a continuing challenge
in modern astronomy.  A number of packages performing point source
extraction has been developed, among them SExtractor
\citep{Sextractor}, StarFinder \citep{StarFinder}, DAOPHOT
\citep{DAOPHOT}, and DoPHOT \citep{DoPHOT}.
 
Due to the specific nature of {\it Spitzer } data and the data collection
strategy there was a need to develop a point source extractor that
would combine the best features of existing programs and extend their
capabilities.  {\it Spitzer } data range from undersampled IRAC
(InfraRed Array Camera) data with very low background to high
background nearly Nyquist-sampled low noise MIPS24 (Multiband Imaging
Photometer at 24 $\micron$) data or very noisy MIPS70 (Multiband
Imaging Photometer at 70 $\micron$) data.  Also for each instrument
different observing strategies result in images that are background
limited or confusion limited.  The point source extractor has to be very
flexible to accomodate this variety of data.

MOPEX has been designed to be applicable in all these cases.
Point
source extraction is implemented as a two step process: point source
detection and profile fitting. 
  MOPEX has
two modes of point source extraction - single frame and multiframe.
In the single frame mode point source detection and subsequent fitting
is performed in the same image. 
In the multiframe mode point source
extraction is performed in a set of input frames.
However, the detection is performed in the mosaic image created by combining the set
of input frames into a single mosaic image. 
The signal-to-noise ratio is
higher in a mosaic image and this fact justifies performing detection
there.  Since mosaicking is part of MOPEX\citep{MOPEX}, point source extraction
benefits from such capabilities as creating properly resampled mosaic
images, cosmic ray hits masking, etc. 
The difference between the single frame and multiframe modes is 
the way the point source fitting is performed.
In the multiframe mode it is done simultaneously in all the input frames. 
The mosaic mode of operation is better suited for well sampled data
where one can obtain a good estimate of PRF in the mosaic image.
Also one should use the mosaic mode for data with high depth of coverage,
since simultaneous point source fitting in a great number of images
becomes computationally prohibitive. 

Point source fitting requires a point response function (PRF).
MOPEX can use a PRF produced by some outside means, but for better
performance the PRF should be estimated from the data itself. MOPEX
has such capabilities.  The package provides means of estimating PRF
for both single and multiframe point source extraction modes.

This paper deals with the single frame (mosaic) mode of point source
extraction and PRF estimation.  Description of the multiframe mode of
MOPEX will be given elsewhere.  The structure of this paper is as
follows. In Section~\ref{ProcessingOverview} we give an overview of
the processing in the mosaic mode.  Background subtraction and noise
estimation are described in Section~\ref{Background}.  In
Section~\ref{Detection} we describe the non-linear matched filtering
technique used to enhance point sources.  We also give a short
description of the image segmentation.  In Section~\ref{Extraction} we
describe the point source fitting performed using a modified Simplex
algorithm.  Passive and active deblending are also discussed there.
Section~\ref{PRFEstimation} is a brief description of PRF estimation
in a mosaic image.  In Section~\ref{Real} we present applications of
MOPEX to two {\it Spitzer} mosaic images and to a Digital Sky Survey (DSS) 
image.  In Section~\ref{Simulated}
we present the results of validating MOPEX with simulated data.
Appealing features of MOPEX are processing speed and photometric
accuracy.  See Section~\ref{Benchmarking} for the timing results of
MOPEX.

\section{Processing Overview}\label{ProcessingOverview}

\begin{figure}
\figurenum{1}
\epsscale{1.0}
\plotone{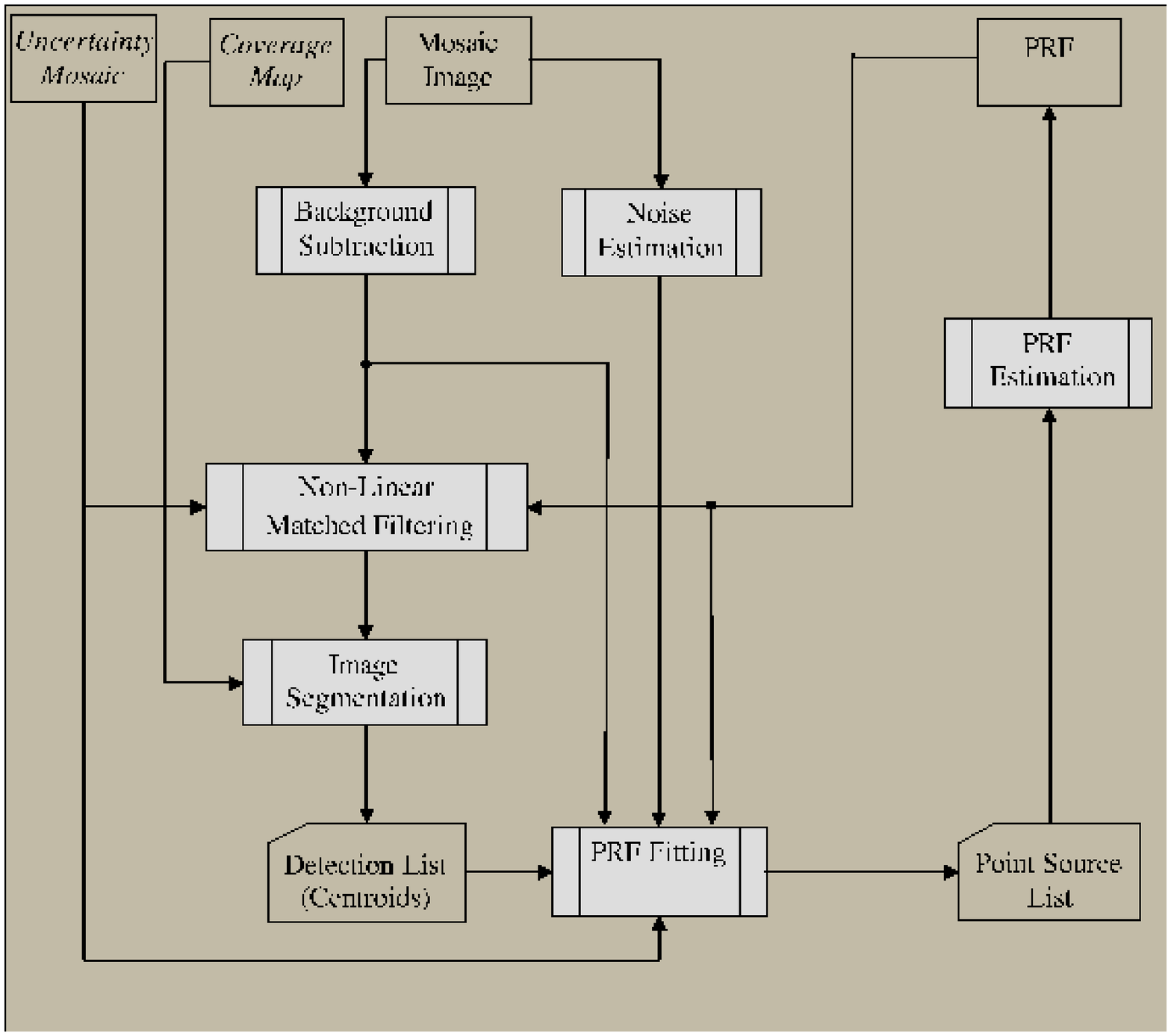}
\vspace{0mm}
\caption{Point source extraction processing chain. The gray boxes represent
the processing steps and the white boxes represent the data. Optional input is
italicized.}
\label{ProcessingChain}
\end{figure}

The mosaic mode of operation is shown in Figure~\ref{ProcessingChain}.
In addition to the mosaic image there are two optional input images:
the coverage map and uncertainty image. The coverage map gives the
number of input frames that were combined to produce each mosaic pixel.
 The uncertainty image gives the uncertainty for each input
image pixel. These two images are created when mosaicking is done with
MOPEX. They are used at a various stages in the point extraction
process. The coverage map is used in point source detection.  The
uncertainty image is used for filtering and point source fitting.
Normally, {\it Spitzer} data come with uncertainty images from which a
mosaic of uncertainty images can be created. If no uncertainty images
exist they can be estimated by MOPEX using the model consisting of three
components - photon noise, read noise, and confusion noise.

Important steps in the processing are creating background subtracted and noise images.
Background subtraction is done by computing the median and subtracting
it.  The background subtracted image is used as an input for the
filtering as well as point source fitting.  The noise image is  produced for
the purpose of computing the signal-to-noise ratio (SNR) of the point
sources.  Optionally, this noise image can be used instead of the
uncertainty image, if the latter is not avalaible and can not be
reliably computed. An efficient sliding window technique is implemented for 
computing both products.
 
For better results filtering is done 
as a preliminary step before image segmentation. 
Filtering can be a simple median subtraction or a more complicated
non-linear matched filtering producing a point source probability image.
Non-linear
matched filtering significantly increases the SNR of the point sources
in the filtered image and also supresses contribution of cosmic ray hits
and various artifacts.  Only in rare cases of images with very low,
almost negligible, noise level and high point source density using
the product of the non-linear filtering for image segmentation can be detrimental.

The result of image segmentation is the detection list with
the positions of candidate point source
are determined.
During the estimation stage a thorough fit of the data is performed to
refine the positions, determine the fluxes and deblend extracted
point sources. The background subtracted image is normally used for the point
source fitting.  A point source subtracted image can be produced to
assess the quality of point source extraction.  A point response
function (PRF) is required to perform linear matched filtering and
data fitting.  MOPEX included a separate task of PRF estimation.
We illustrate the intermediate products at various processing
stages in Figure~\ref{Illustration} using a small fragment
(90 by 60 pixels) of an MIPS70
mosaic image. The mosaic is described
in more details in Section~\ref{RealMIPS70}.  

In this paper we present an overview of
the algorithm.  One can create a custom-tailored
processing chain to accomodate for any specific features of the data,
as descrived, for example, in Section~\ref{RealMIPS24}.
Careful tuning of various stages
can be performed using a number of parameters.
 For details on running MOPEX, comprehensive
listing of all the processing parameters, and sample namelists one
can consult the online guide
\footnote{http://ssc.spitzer.caltech.edu/postbcd/}.

\section{Background and Noise Estimation}\label{Background}
Background subtracted images are used for both point source detection
and fitting.  A common way of estimating the background is to find for
each pixel the median in a window centered around that pixel. This
method will inevitably overestimate the background in the vicinity of
bright sources. MOPEX provides a user controlled way to alleviate the
bias introduced by the bright sources.  One can counter-bias the
median by excluding a fixed number of pixels with highest values in
each window from median computation, thus offsetting the bias
introduced by the presence of bright sources.
 This  number is defined
by the user and is constant througout the image to which it is applied.
This step may require further optimization by automatically adjusting
this  number based on the crowdiness in any particular window.
However, even in its present state the quality of median subtracted
 images satisfies the needs of point source detection. 
Any bias in the background
estimation introduced by the median filtering can be
corrected at the point source fitting stage as described below 
in Section  ~\ref{Extraction},  since the fitting includes among other fitting
parameters a constant background for the fitting area.

By increasing the window size one can decrease the fluctuations
in the number of point sources per window and in general get a more
robust estimate of the median. The problem with increasing the window
size is the corresponding increase of the processing time.  In order
to find the median the values of the pixels in the window should be
sorted.  If sorting is done from scratch for each window this process
becomes prohibitively slow.  To speed up the process and make median
filtering practical for relatively big windows, we use the sliding
window approach.  The pixel values are sorted in the first window
which is located in a corner of the image.  Subsequently, as the
window slides by one column the pixels from the dropped out column are
removed from the sorted list of pixels and the new column is inserted
in the sorted list of pixels.  
Various data structures have been used for fast median computation
(see \citet{median} and references therein).
 We implemented a variant of the binary search tree method.
This approach significantly speeds up
the processing.  The processing time scales with the linear size of
the sliding window, whereas if the sorting is performed from scratch
in each window the processing time scales at least as the area of the
window.  On a 1 GHz Sparc Sun workstation processing time for
computing the median for a window of size 100x100 pixels is $10^{-4}$
sec/per pixel.
  
Noise images are used for computing the  SNR of
the point sources.  Noise images for each pixel give a value of
background fluctuation in a window centered on the pixel. They are
computed based on the assumption of Gaussian distribution of pixel
values around the underlying sky value.  The presence of point
sources results in broadening of the pixel distribution and
overestimation of the noise level. Just like in the background
estimation this can be alleviated by excluding the highest pixel
values in each window.  Noise estimation is an extension of median
filtering.  The median pixel $i_{med}$ is found for each window.  Then
the noise value $n$ is determined as
\begin{equation}
n = (w[i_{med} + 0.34 W] -  w[i_{med} - 0.34 W])/2.
\end{equation}
Here $w$ is the sorted array of pixels in the window, $W$ is the size
of the array.  The same sliding window technique is used here as in
background estimation.

\section{Point Source Detection}\label{Detection}

Point source detection is performed either on background subtracted
images, or optionally additional filtering can be performed.

\subsection{Non-linear Matched Filtering} 
The purpose of filtering is to reduce fluctuations of the background
noise and to enhance point source contributions.  It is a common
practice to perform linear matched filtering to reach the above goals
\citep{Andrews}.  A matched filter can be derived \citep{Cook} on the
basis of optimizing SNR, likelihood ratio, or mean square error (MSE).
Standard derivation of linear matched filter involves an assumption of
either a single point source or the Gaussian distribution of the point
sources.  In reality the distribution function of point sources is
highly non-Gaussian.  Under such condition the linear filter becomes
sub-optimal and the optimal filter is non-linear. The general form of
a such a filter is very complicated ~\citep{SPIE} and its derivation
involves fitting the point source distribution function with the
Gaussian mixture model. Application of such a filter is
computation-intensive.  For practical purposes we take the general idea of
non-linear filtering and derive a non-linear filter based on the
notion of point source probability.  The details of the derivation are
given in Appendix~\ref{app_matched}.  A point source probability image
is computed as follows:
\begin{equation}\label{PSP_equation}
PSP(j) = (1 + \frac{(1-Pr) \sigma_x } {Pr \sigma_T} exp(-\frac {
  (\sum_i s(i) PRF(i) )^2} {2\sigma_n^4/\sigma_T^2} ) )^{-1},
\end{equation}
where $s(i)$ is the pixel value of the input image,
$PSP(j)$ is the value of pixel $j$ in the point source
probability image and $Pr$ is an apriori probability of point source
presence.  Quantities $\sigma_x$, $\sigma_n$, and $\sigma_T$ are
defined in Appendix~\ref{app_matched}.  The algorithm is not sensitive
to the exact value of $Pr$ which is set by default to 0.1.
Figure~\ref{Illustration}b shows the $PSP$ image corresponding to the
input image in Figure~\ref{Illustration}a.  Convolution with the PRF
causes some smearing of the point sources.  The smearing,
however, is not of concern, since the filtered images are used for
detection only.

\subsection{Image Segmentation}

\begin{figure}
\figurenum{3}
\plotone{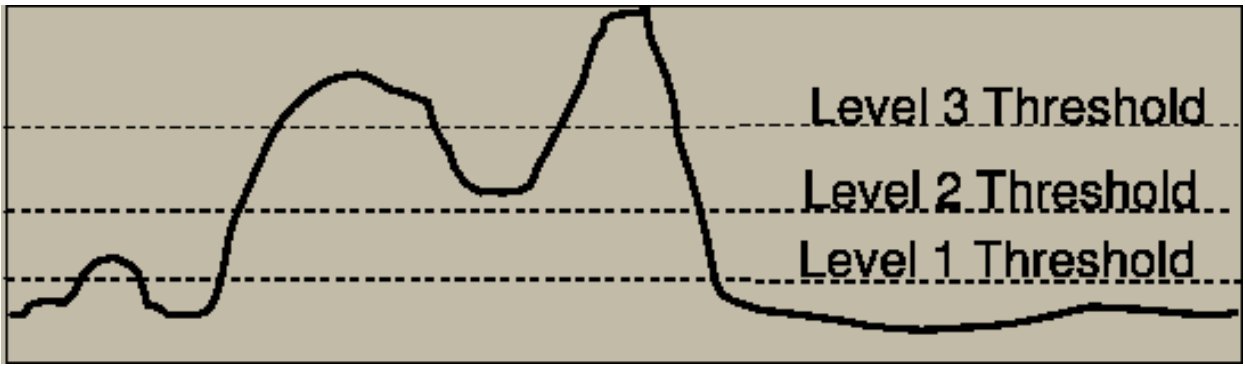}
\vspace{0mm}
\caption{Shown here is a cross-section of an image being segmented.
Pixels with the values higher than a segmentation threshold 
are grouped into contiguous clusters. 
The segmentation threshold is raised in order to reduce the sizes of the
detected clusters of pixels.}
\label{Threshold}
\end{figure}

Filtered images undergo a process of image segmentation. Contiguous
clusters of pixels with the values greater than a user-specified
threshold are identified.  If the number of pixels in a cluster is smaller than a
user-specified threshold, the cluster is rejected. This procedure
serves as an additional guard against cosmic rays affected pixels and
peaks in background noise fluctuations.  If the number of pixels is
greater than a user-specified threshold, the cluster is subjected to
further segmentation with a higher threshold (Figure~\ref{Threshold}).

The process of raising threshold is very sensitive.
Initially the threshold $T$  defined in terms of a
robust estimate of the background fluctuations in the filtered
image. The subsequent increase of the threshold is
is  the essential part of passive de-blending as described below 
in Section~\ref{Deblending}.
The simple approach implemented in MOPEX is as follows.
For each cluster a new value of the threshold is found
based on the mean pixel value and the standard deviation
of the pixel values in the cluster. The drawback of 
such approach is that it misses a lot of sources
which are close to brigher sources. It the original 
clusters contains one bright and several faint sources
the raised threshold will miss the faint sources. 

A more complicated scheme has been implemented
to facilitate passive de-blending. This scheme
uses the concept of {\it peak} pixels. A pixel is declared a {\it peak} pixel
it its value is greater than the values of a user specified number of 
adjacent pixels. In this scheme the cluster is split
as long as there are more than one {\it peak} pixel in it.
The minimum  {\it peak} pixel value in the cluster $P_{min}$
is determined. The effective segmentation threshold $T_n$
for the detection level $l$ is equal to 
\begin{equation}
T_n = P_{min} - \frac{T}{l}.
\end{equation}
This scheme ensures that no faint sources are missed in the
presence of a bright neigbor.

\begin{figure}
\figurenum{2}
\plotone{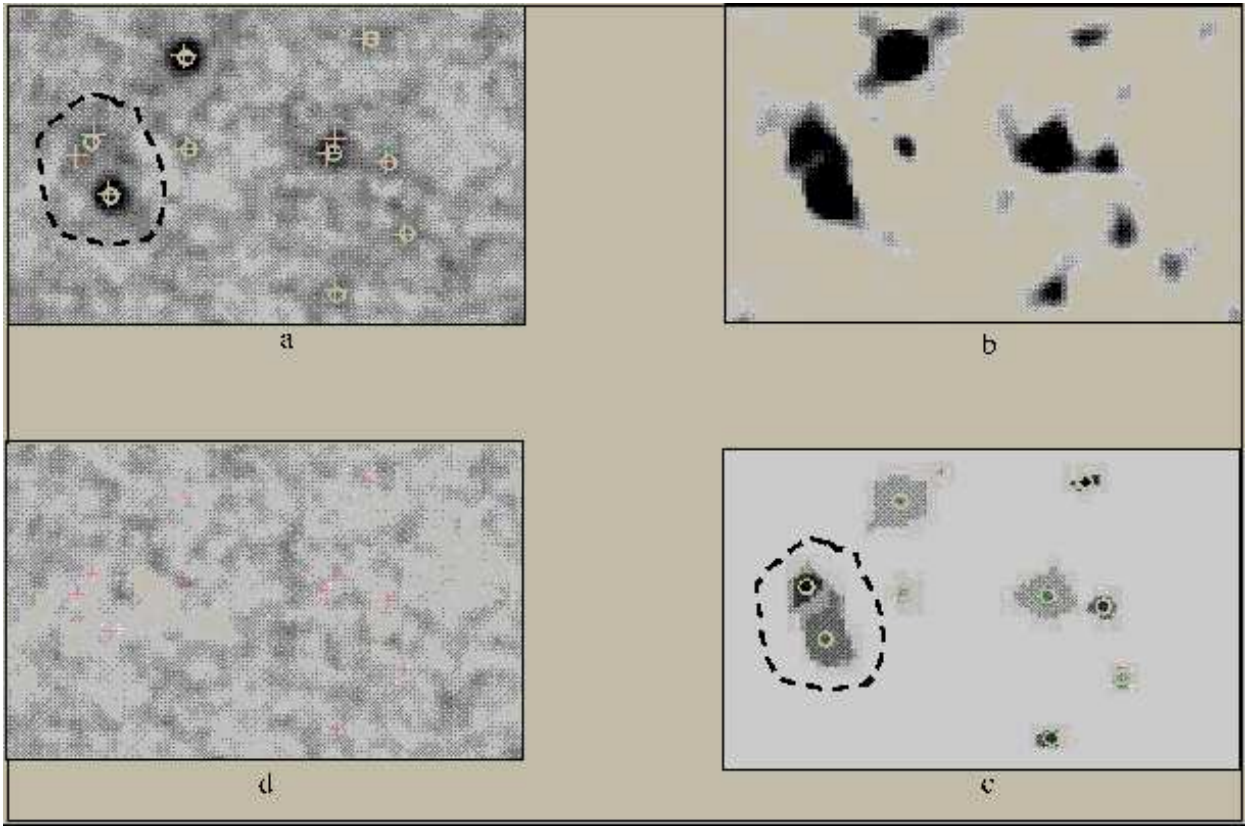}
\vspace{0mm}
\caption{A small fragment (90 by 60 pixels) of MIPS70 mosaic shown
 in Figure~\ref{MIPS70} is used to illustrate the products
of  the processing chain. a)
Input image with the detections shown as {\it white circles}, and
final extractions shown as {\it white crosses}. An example of active
deblending is circled with the {\it black dashed line}. b) Point source
probability image. c) Detection map with the detections shown as {\it
white circles}. An example of a blend with $N_p$ = 2 is circled with a
{\it black dashed line}.  The {\it white dashed line} circles
 the detection that was discarded because of its small size.  
d) Point source subtracted image.}
\label{Illustration}
\end{figure}

The program keeps track of threshold raising and assigns a detection
level to each pixel based on the last threshold value for which this
pixel was above the threshold.  The original threshold corresponds to
the detection level 1.  Each time the threshold is raised the
detection level is incremented.  A detection map image can be created
to visualize the image segmentation process.
Figure~\ref{Illustration}c shows the detection map corresponding to
the input image in Figure~\ref{Illustration}a. It has an example of
a detection that was discarded because its size was smaller than
the user-specified threshold; it is circled with the white dashed line.
Higher levels of
detection are shown with darker shades of gray.  When the process of
segmentation is finished the centroids of the clusters are calculated
and stored in the detection list (see
Figure~\ref{ProcessingChain}.)  Centroids that belong to the same
first level detection clusters are marked in the detection table as
part of the same blend.  The size of the blend defines the initial
$N_{p}$ for point source fitting below in equation~\ref{chi2}.  An
example of such a blend with $N_{p} = 2$ is circled in
Figure~\ref{Illustration}c with a black dashed line.

Mosaic images in general have variable coverage. 
Applying a constant pixel value threshold results in variable effective
threshold; areas of higher coverage will have higher threshold in
terms of the SNR of the detected point sources. To overcome this
problem the coverage map is used to attenuate the image undergoing
segmentation.  The input image is multiplied by the square root of the
coverage before it undergoes the process of segmentation. 
It can be done for any input image, i.e. either the background subtracted
image or the probability map. 
 Also, optionally
one can specify minimum coverage to prevent detections in the
areas of low coverage, which are usually very noisy and where
detection can not be performed reliably even after the filtering is
performed.

\section{Point Source Fitting}\label{Extraction}
\begin{figure}
\figurenum{4}
\plotone{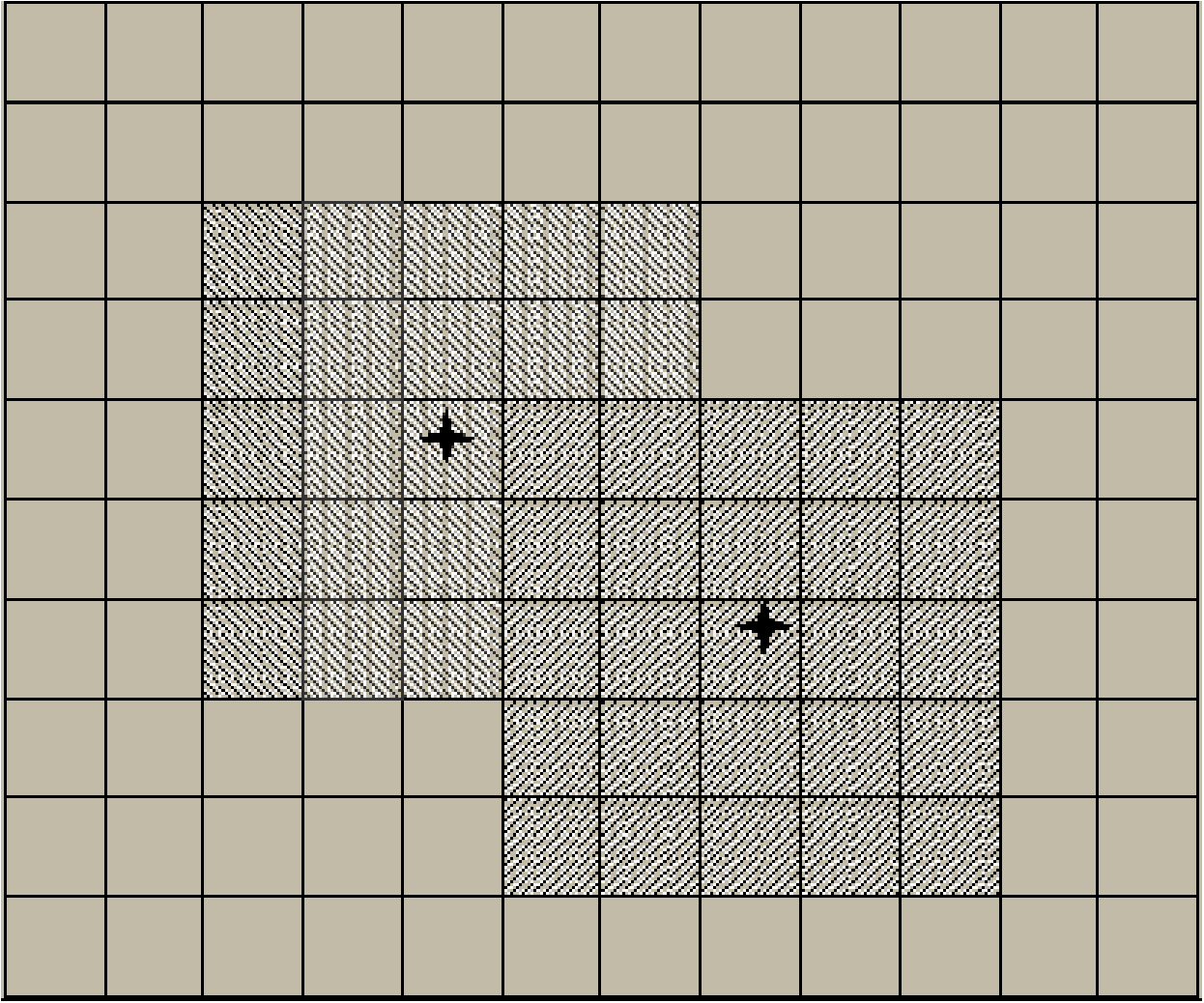}
\vspace{0mm}
\caption{Fitting area for simultaneous fitting of two point sources.
The detection positions are shown with {\it black crosses}. Each
detection has a fitting area of a 5 $\times$ 5 pixels square.}
\label{FittingArea}
\end{figure}

Final point source position and photometry estimation is performed for
all detections on the detection list. For each point source
candidate the data in the input image is fit with the PRF.  Fitting is
performed by minimizing $\chi^2$:

\begin{equation}\label{chi2}
\chi^2 = \sum_{i\in W}\frac{ (s(i) - \sum_{n=1}^{N_p} f_n PRF(i,{\mathbf
R}_n) - b)^2}{\sigma^2(i)}.
\end{equation}
Here the summation is performed over pixels $i$ from the fitting area
$W$; $s(i)$ and $\sigma(i)$ are the pixel values of the input image
and uncertainty image, correspondingly;  $f_n$ and ${\mathbf R}_n$ are
the flux and the position of the $n-th$ point source; $PRF(i,{\mathbf
R}_n$) is the contribution of the $n-th$ point source to the $i-th$
pixel; $b$ is the constant background within the fitting area that can
be used in this formula optionally. 
The number $N_p$  of point sources  fit simultaneously is
set to 1 initially if the detection does not belong to a blend,
 as described in the previous section.  For the detections belonging to a blend
 the number $N_p$ is initially set to the size of the blend.
  PRF contribution is computed
for any fractional position of the point source; a bilinear
interpolation is performed from the grid points available in the PRF.
    The fitting area $W$ is a combination
of rectangle areas centered on the detection positions (Figure
~\ref{FittingArea}).  The size of each rectangle is specified by the
user and should be set to be on the order of the size of the Airy disk. 
If set properly, the fitting areas of the point sources belonging
to one blend are partially overlapping. As a result of detection
  point sources with the overlapping contributions should
  belong to one blend. However, if the detection is done poorly,
  e.g. if the detection threshold is set
  too high, then such sources will be erroneously 
  put in two separate blends and not fit simultaneously.
 If the uncertainty image is not available,
the noise image can used instead. 
Users have an option of using the background subtracted
mosaic, using the original image and fitting the background for each point source, 
or doing both.

The  strategy used for minimization of $\chi^2$ 
 is a hybrid of a modified simplex algorithm and 
the gradient descent algorithm.
The simplex algorithm does not use derivatives of the
functions involved in minimization.  It is a desirable feature since
the world coordinate transformations are very complicated
functions of their arguments. We made several modifications
of the algorithm, which are described in Appendix~\ref{app_simplex}.
Simplex operations are applied to the point sources positions.
On the other hand the derivatives with respect to the fluxes of the point sources
and the background are taken easily:
\begin{eqnarray}\label{dchi2_dflux}
\frac{\partial \chi^2}{\partial f_k} = -2\sum_{i\in W}\frac{ PRF(i,{\mathbf R}_k) (s(i) - \sum_{n=1}^{N_p} 
f_n PRF(i,{\mathbf R}_n) - b)}{\sigma^2(i)};\nonumber \\
\frac{\partial \chi^2}{\partial b} = -2\sum_{i\in W}\frac{ (s(i) - \sum_{n=1}^{N_p} 
f_n PRF(i,{\mathbf R}_n) - b)}{\sigma^2(i)};
\end{eqnarray}
 and that justifies the choice of the gradient
descent method for the fluxes and the background.

In general, several point sources are used to fit the data
simultaneously as explained in the next section.
A $2-$dimensional simplex is constructed for the position vector
of each point source. The appropriate simplex operations is performed for each
point source separately.
After the positions of the point sources are adjusted
each flux and the background are adjusted along the gradient given by the equation~\ref{dchi2_dflux}.
The procedure is repeated until one of the stopping criteria specified below is met.
Since each simplex is moved separately, it reduces the complexity
of the algorithm. The reflections, etc. are performed in
the $2-$dimensional space of the position vertor 
for each point source, as opposed to the $2N_p$ dimensional
space of the vector describing the position of all $N_p$ sources.

 The goodness-of-fit is assessed by the
values of $\chi^2/dof$.  The number of the degrees of freedom $dof$ is
equal to the size of the fitting area $W$ minus the number of the
fitting parameters, which is 3 per point source and one for the background,
if it is used in the minimization.
The program attempts to minimize $\chi^2/dof$ to be below the user
specified threshold. If the number of iterations exceeds the limit or
the relative change of $\chi^2$ drops below the limit, fitting
terminates, even though the $\chi^2$ is still greater than the
threshold.

\subsection{Passive and Active Deblending}\label{Deblending}
If several point sources are within the reach of each others
PRF's they should be fit simultaneously. This is done if $N_p>1$
in equation~\ref{chi2}.
This process is known as point source deblending.
There are two mechanisms in MOPEX to perform deblending. 

The first one is called passive deblending.
The point sources identified at the detection stage as belonging to one
blend are fit simultaneously. In this case fitting starts
with the value $N_p>1$.

The second mechanism is called active deblending.
During active deblending $N_p$ is incremented during the fitting process.
If $\chi^2/dof$ for a point source is
above the user specified threshold, the algorithm increments $N_p$ and
fits the same data with more point sources. If the improvement in
$\chi^2/dof$ is significant, then the additional point source is
accepted. Otherwise the algorithm reverts to the previous solution. In
case of succesful active deblending if $\chi^2/dof$ is still greater than
the threshold, active deblending continues untill the program reaches
the user defined limit on $N_p$. 

\begin{figure}
\figurenum{5}
\plotone{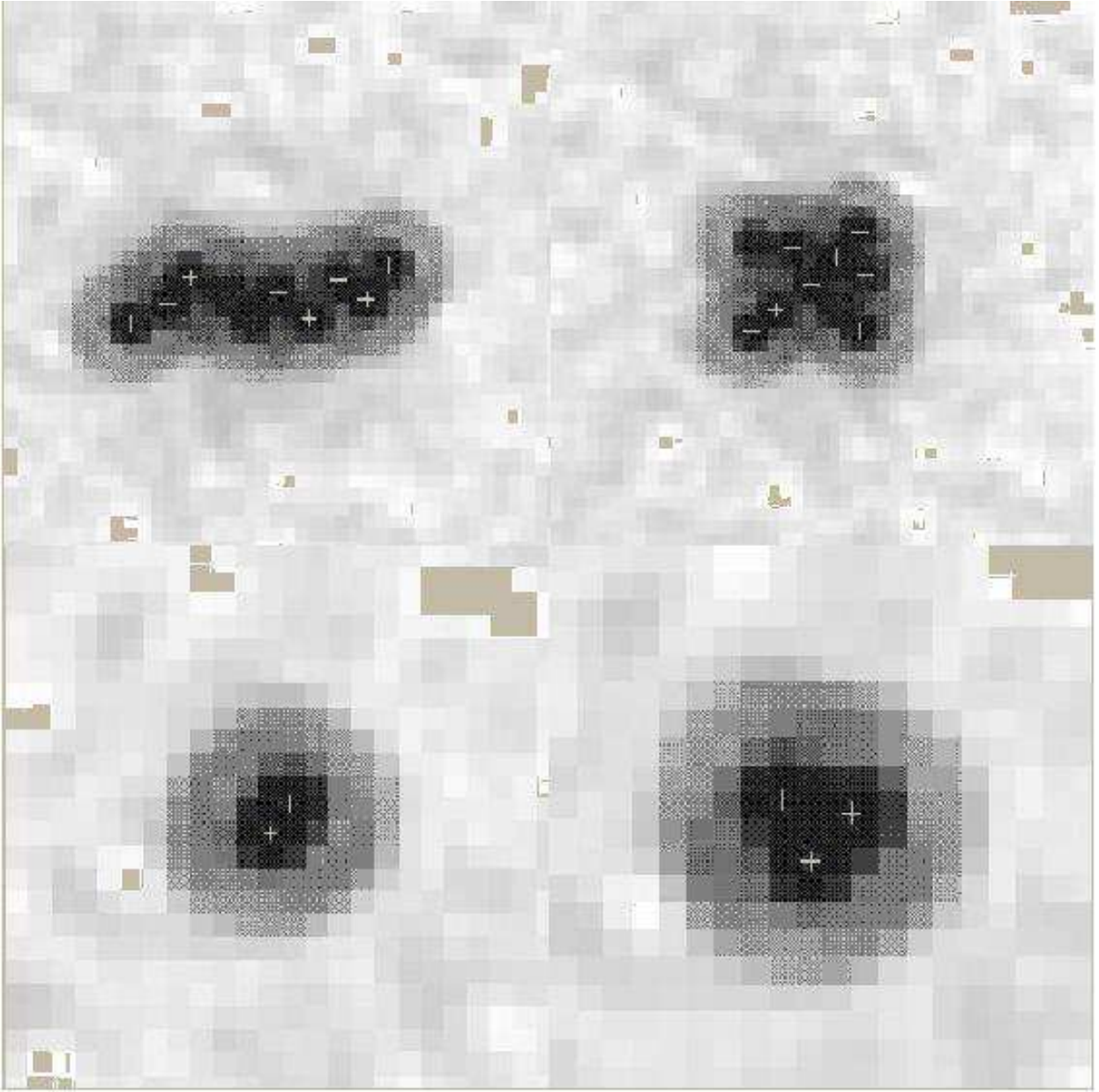}
\vspace{0mm}
\caption{Shown here are four simulated images used for testing of the 
passive (top row) and active (bottom row) deblending.
The sources in the top row are separated by at least one FWHM of
 the PRF, the 2 sources in the bottom row left picture are
separated by $\sim$ 1/2 of the FWHM, and the 3 sources in the 
bottom row right picture are separated by $\sim$ 2/3 of the FWHM.
The locations of the source are shown with  {\it white crosses}.}
\label{MIPS_Sim}
\end{figure}

We performed simulations to test the limits of passive 
and active deblending. To test passive deblending
a set of point sources with the same flux was added to a smooth background
image. The average separation between the adjacent sources was
approximately the FWHM (full width half maximum) of the PRF.
Two examples of such a cluster with 10 point source is shown in
Figure~\ref{MIPS_Sim}. These sources were separated at the detection stage,
i.e. they were detected as separate sources.
They were marked as belonging to one initial cluster and were
fit simultaneously. 
The simulation was repeated several times for a number of 
sources ranging from 2 to 20.
In Table~\ref{MIPS_Sim_P} we show the processing time and
the accuracy of determining positions and fluxes.
The positional error $\delta_R$ is defined as the
average distance between the true ${\mathbf R}_n^{T}$
 and extracted ${\mathbf R}_n^{E}$
positions of the point sources $n$:
\begin{equation}\label{deltaR}
\delta_R = \frac{1}{N_p} \sum_n^{N_p}|{\mathbf R}_n^{T} - {\mathbf R}_n^{E}|.
\end{equation}
The relative flux error $\delta_f$ is defined as 
the average ratio of the absolute difference between 
the true $f_n^T$ and extracted flux $f_n^E$ to the true flux:
\begin{equation}\label{deltaf}
\delta_f = \frac{1}{N_p} \sum_n^{N_p}\frac{|f_n^{T} - f_n{E}|}{f_n^{T}}.
\end{equation}
The processing time
was measured on a 1 GHz Sparc Sun workstation 
with the fitting area of the size of 7 by 7 pixels.
The processing time depends on a variety of other
factors, so the numbers quoted in Table~\ref{MIPS_Sim_P}
can be used as a general guide only.

To test active deblending we performed similar simulations.
The goal of these simulations was to test the limit 
in terms of the algorithm's ability to de-blend sources
that cannot be separated at the detection stage.
We simulated images with the clusters of 2 and 3
point sources. The distance between the sources 
in the 2-source clusters was $\sim$ 1/2 of the 
FWHM, the distance between the sources in the 3-source
clusters was $\sim$ 2/3 of the FWHM.
In the bottow row of Figure~\ref{MIPS_Sim} we show an example 
of a 2-source cluster and a 3-source cluster.
The positional and flux errors were
$\delta_R = 0.03$ pixel and $\delta_f = 0.01$ for the
2-source clusters and $\delta_R = 0.01$ pixel and $\delta_f = 0.004$
for the 3-source clusters. The numbers are slightly
lower for the 3-source clusters. This is explained
by the fact that the separation is greater for 
these clusters. The important figure of merit
is the failure rate $ER$, which is the fraction
of the cases for which the algorithm
could not succesfully deblend the sources.
For the 2-source clusters $ER \sim 0.06$.
For the 3-source clusters, even though
the separation is greater, the failure rate
increases dramatically to $ER \sim 0.25$.
Active deblending of clusters with more than 3 sources
is not reliable.

An example of passive and active deblending in the real data 
is given in Figure~\ref{Illustration}a.
The cluster of sources for which both passive and active deblending
was performed is circled with the dashed line.

\subsection{Output} 
The output is a list of point sources that specifies their world and
local coordinates, fluxes, uncertainties, goodness-of-fit measure,
estimated SNR, and a number of other quantities that are used for
quality assessment.  The flux, position and background uncertanties $\sigma_{z_i z_i}$
and cross-correlation $\sigma_{z_i z_j} $ are determined
by the diagonal elements of the inverse of the Hessian matrix:
\begin{eqnarray}
H_{z_i z_j} = \frac{\delta^2 \chi^2}{\delta z_1 \delta z_2};
\label{hessian}
\end{eqnarray}
where $z_i, z_j$ are the fit parameters $f$, $x$, $y$, and $b$.
\begin{eqnarray}
\sigma_{z_i z_j} = H^{-1}_{z_i z_j};
\label{hessian_errors}
\end{eqnarray}

SNR for point source with flux $f$ and position ${\mathbf R}$ is
computed as
\begin{equation}
SNR = \frac{f}{N({\mathbf R}) NP}
\label{SNR_definition}
\end{equation}
where $N$ is the noise image, described in Section~\ref{Background}; $NP$
is defined as the effective number of pixels whose noise contributes
to the measurement of the flux of the point source.

One can specify an unlimited number of apertures to compute aperture
photometry for each extracted point source.

\subsection{Software} 
MOPEX consists of a number of modules written in C/C++ which are glued
together by perl scripts. Specifically, the point source extraction in
the mosaic mode discussed in this paper is performed by the perl
script {\sl apex\_1frame.pl}.  PRF estimation is performed by the perl
script {\sl prf\_estimate.pl}. Point source subtracted images
are created with the perl script {\sl apex\_qa.pl}. The software parameters are input through namelists,
ASCII configuration files. The software is available 
for download at the {\it Spitzer} website\footnote{http://ssc.spitzer.caltech.edu/postbcd/download-mopex.html}.
The documentation included in the distribution has a detailed
description of the parameters used in point source extraction
as well as sample namelists.

\section{PRF Estimation}\label{PRFEstimation}
Here we give a brief description of PRF estimation, which will be
described in more details elsewhere.  The first time point source
extraction can be performed with a theoretical PRF or even a Gaussian
PRF.  It is performed with a high detection threshold in order to find
bright sources. An additional selection should be performed to find
only non-confused point sources. A set of postage stamp images is cut
out from the background subtracted mosaic image centered on each 
point source position. More accurate point source positions are
estimated by fitting a Gaussian to the postage stamp images.  There is
an option of rejecting and replacing outlier pixels and outright
rejecting "bad" images.  The postage stamp images are resampled and
shifted using bicubic interpolation \citep{bicubic} and combined into
one final PRF image.

We want to emphasize here that the use of the term PRF is not
just an alternative way of saying point spread function (PSF). 
PSF-fitting is a commonly used term. However, PRF and PSF
are two different objects. PSF is an image a point source.
PSF is often oversampled, i.e. the pixel size of the PSF image
is a fraction of the pixel size of the detector array or the 
mosaic image for which the PSF is applicable. 
 PRF, however, is {\bf not} an image of a point source.
It is a table of values of responses of the detector array (or mosaic)
 pixels to a point source. The positions of the pixel for which the
response is calculated are on a grid. Normally the PRF is 
oversampled, which means that the spacing of the grid is 
a fraction of the detector pixel size. 
The two - PRF and PSF - are one and the same if they are not oversampled, i.e.
the pixel size of the PSF is equal to the pixel size of 
the detector array(mosaic) and equal to the grid spacing
of the PRF. 
The advantage of using PRF vs. PSF has to do with the way
they are derived from the data. PRF is closer to what
is being observed by the detector array pixels. 
Every processing step inevitable introduces 
errors in the product being calculated.
The only error introduced in estimating PRF 
is the one produced by the shift of the observed
pixels to the fixed grid. Estimation of the PSF involves
an additional step of interpolating from the bigger
detector array pixels to the smaller PSF pixels.
This step introduces an additional error. 
However, for the purposes of fitting one has to reverse this
  and integrate the small pixels back into the bigger
 detector array pixels.

\section{Application to {\it Spitzer} and DSS data}\label{Real}

We performed point source extraction on the mosaic images taken at 24
and 70 $\mu$m with MIPS as part of the extragalactic First Look Survey (FLS)
\citep{Fadda,Frayer}. These images consist of a main shallow survey,
centered on a region near the ecliptic pole (RA[J2000]=17:18,
Dec[J2000]=+59:30) and covering 4.4 square degrees, and deeper
observations (verification strip) contained in the main survey but
smaller in size (0.26 square degrees).

\subsection{MIPS24 Mosaic}\label{RealMIPS24}
The 24 $\mu$m data was mosaicked to include both the shallow (median
coverage of 23) survey and the smaller verification strip (median
coverage of 116). The mosaic was created with square pixels measuring
half of the detector's original pixel size, i.e. 1.275~arcsec. The
effective integration time for the verification strip is $\sim$426
seconds, about five times deeper than the main field with its $\sim$84
seconds. The noise at 3$\sigma$ is 0.08 mJy (verification strip) and
0.16 mJy (main field).

Point source extraction is performed on the mosaic image, since MIPS24
data is well sampled and usually have a high coverage depth as is the
case for the FLS data.  Detection and point source fitting is done in
a 5 step process. The first step consists of selecting point sources
for PRF estimation.  As an initial guess a theoretical PRF produced by
the {\it Spitzer} version of the Tiny Tim Point Spread Function (PSF)
modeling program\footnote{Developed for the {\it Spitzer} Science
Center by John Krist; STScI} was used.  Point source extraction is
performed without doing active deblending and only the brightest
non-confused (non belonging to a blend of detections) sources (flux $>
5 \times 10^3$ microJy) with the lowest $\chi^2$ ($\chi^2/dof < 30$)
are kept. A total of 27 sources are selected for PRF estimation. The
second step is to estimate the PRF based on the selected sources. The
first Airy ring of the PSF is at a radius of 7 mosaic pixels.  We
select a PRF postage stamp size of 35 by 35 pixels and a circular PRF
with radius of 11 pixels (beyond the first Airy ring). The PRF flux is
normalized within that radius and therefore an aperture correction (a
factor of 1.156) needs to be applied to all fluxes \citep{Fadda}.

The first bright Airy ring around the brightest point sources in our
mosaic image are the cause of many false detections in the mosaic
image. Therefore, our third step is to create a mosaic image where all
the point sources with SNR $>$ 20 in the point source probability
image have had their Airy rings removed. This is an example of how the
basic processing chain can be modified in order to accomodate specific
features of the data.  A total of 1224 sources are extracted. Then a
residual image is created where the PRF used for subtracting point
sources has a hole with a radius of 5 image pixels (where the first
minimum occurs in the PRF).

\begin{figure}
\figurenum{6}
\epsscale{0.5}
\plotone{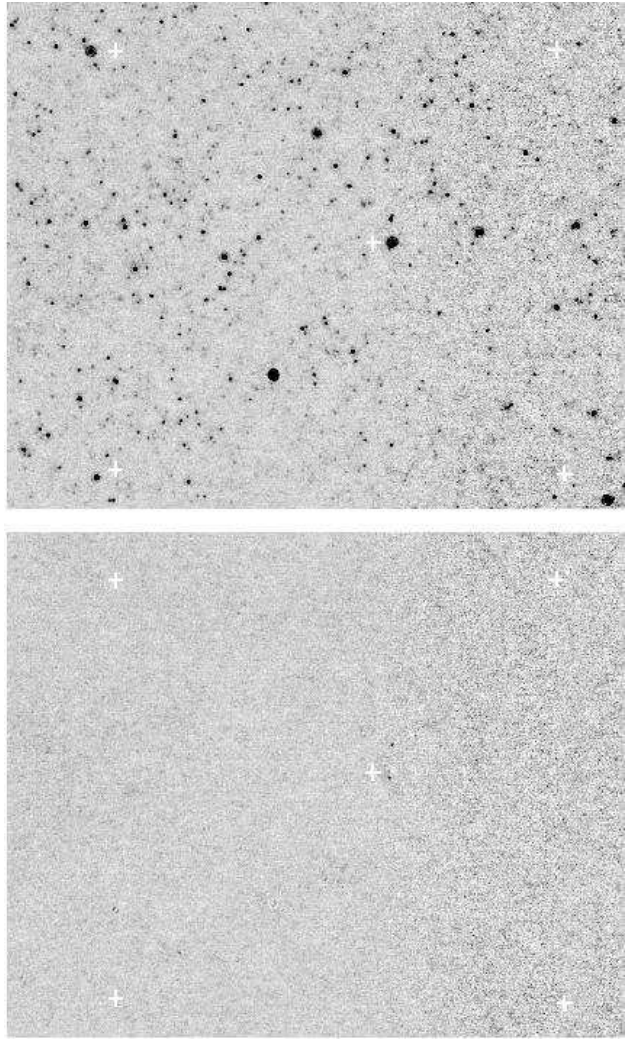}
\vspace{0mm}
\caption{Shown here is a section (1000 by 800 pixels or 0.1 square degree)
 of the MIPS24 FLS mosaic
with the point sources ({\it top}) and after point source subtraction
({\it bottom}). The section is divided between the verification strip
({\it left part of the mosaic}) and the main field outside of the
verification strip ({\it right part of the mosaic}).  Several evenly
spread positions are marked with {\it white crosses} as reference points in
both mosaic images.}
\label{MIPS24}
\end{figure}

The detection is then done on the ringless image with a much lower
detection threshold.  A total of 97512 sources are detected.  The
final fitting is done with passive and active deblending.  A total of
41559 sources ($\sim 1.0 \times 10^4$ sources per square degree) with
SNR $>$ 3 are extracted.  Figure~\ref{MIPS24} shows a section of the
mosaic (1300 by 1000 pixels or 0.16 square degree in size) that has
the verification strip and the main field outside of the verification
strip.  The point source subtracted mosaic is also shown to assess the
quality of point source extraction.

\subsection{MIPS70 Mosaic}\label{RealMIPS70}

The 70 $\mu$m data was mosaicked to contain only the verification
strip (median coverage of 35). The mosaic covers 0.4 square degrees
and the pixel size is 4 arcsec. The effective integration time is
$\sim 210$ seconds. The total number of extracted point sources with
fluxes $ > 3\sigma$ ($\sim 6$ mJy) is $\sim 400$.  The processing
steps here are similar to the ones used for the MIPS24 mosaic.  The
only difference is that we did not create a "ringless" image for the
detection. The reason the Airy rings did not cause any false detection
is that even for the brightest sources they are below the noise level.
The main reason we use the mosaic mode of point source extraction for
MIPS70 data is that in the single frame the data is too noisy to do
any dependable point source fitting and PRF estimation.

\begin{figure}
\figurenum{7}
\epsscale{1.1}
\plotone{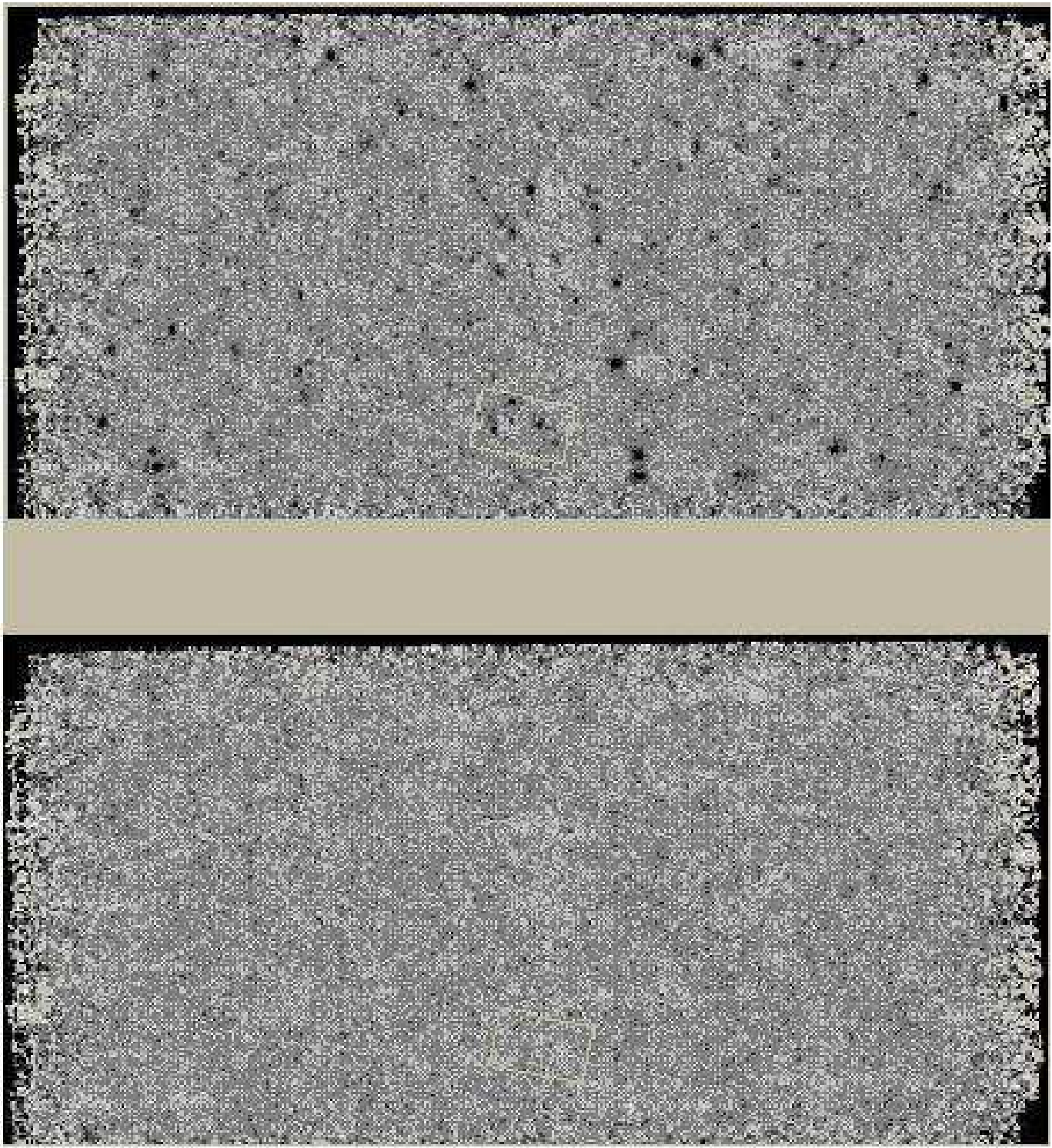}
\vspace{0mm}
\caption{The {\it top image} is the MIPS70 mosaic for the verification
strip of the FLS. The {\it bottom image} is the mosaic with point
sources subtracted. The {\it white rectangle} shows the fragment of
the mosaic used for illustration of processing steps in
Figure~\ref{Illustration}.}
\label{MIPS70}
\end{figure}

Figure~\ref{MIPS70} shows the MIPS70 mosaic of the verification strip
with and without the point sources.  The white rectangle shows the
fragment of the mosaic used for illustration of the processing steps
in Figure~\ref{Illustration}.

\subsection{Digital Sky Survey Image}\label{DSS}

\begin{figure}
\figurenum{8}
\epsscale{0.5}
\plotone{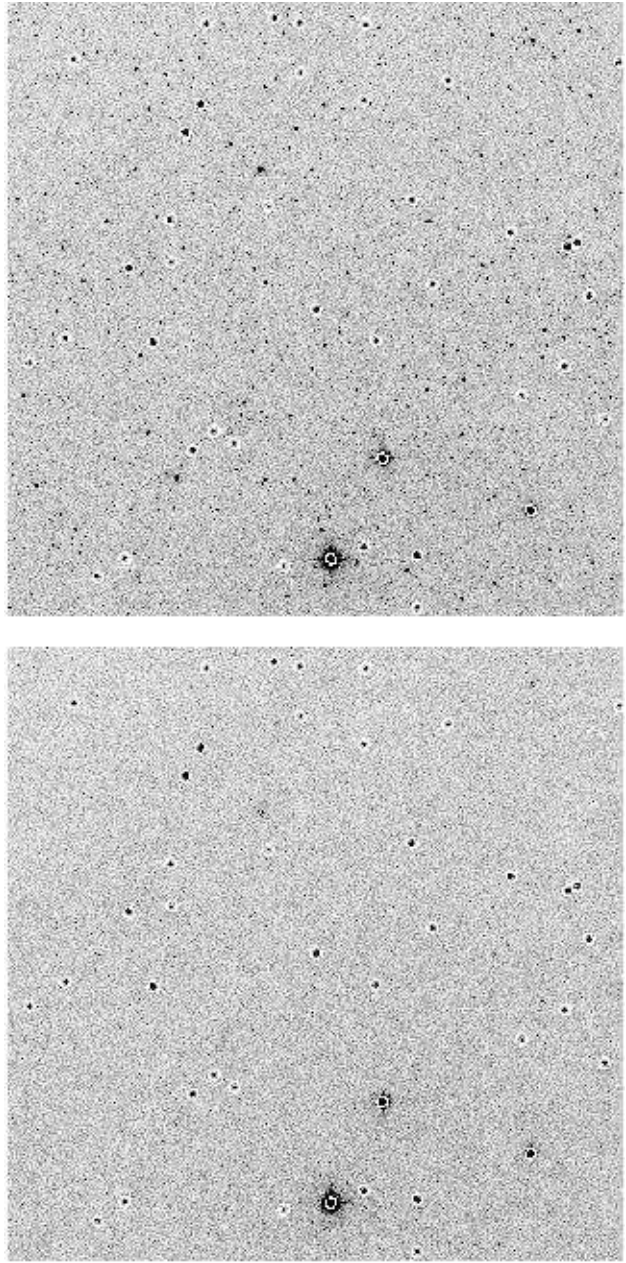}
\vspace{0mm}
\caption{The top image is the DSS image described in the text.
 The bottom image is the same image with point
sources subtracted. The white cirles mark the saturated
or extended sources that could not be fit to any satisfaction. }
\label{DSS_image}
\end{figure}

We also tested MOPEX on a Digital Sky Survey
(DSS\footnote{http://archive.eso.org/dss/dss}) image, of size 40
arcmin across, near the standard star Landolt 92-288 at position
RA[J2000]=00:57:17.093 and Dec[J2000]=+00:36:47.76 (DSS1, plate
dss27753). A total of 2836 sources were detected.  
Figure~\ref{DSS_image} shows
the DSS image and the residual image.
Point source extraction was done in two steps.
A number of sources in the images have their responses
distorted due to non-linear response and several brighter ones
are saturated. Also there is a number of
extended sources in the field. 
The separation between good sources, non-linear sources and 
saturated sources can be done based on the $chi^2/dof$ 
of the fit.
At the first step only 2725 good sources were detected and removed
from the input image. At the second step 111 non-linear sources
were detected and removed from the image. A separate 
PRF was estimated for the non-linear sources.  
The quality of the point source extraction and removal
for the non-linear sources is obviously worse than for 
the good sources, but was deemed satisfactory.
The fitting and removal for the saturated source
cannot be done succesfully. 
These sources are marked in the Figure~\ref{DSS_image} with white circles.

\section{Validation with Simulated Data}\label{Simulated}
MIPS24 mosaics were simulated by adding point sources to the computed
noise images (main and verification). The simulated fluxes were
randomly taken from the flux distribution as given by the number
counts in the main and verifications surveys (Marleau et al. 2004). A
list of true sources was created.  The true sources were convolved
with the PRF derived previously.  Point source extraction done for the
real data was repeated with the same steps and parameters on the
simulated data.

The list of the true sources
used in the simulated images is matched with the list of the extracted
sources. Differential completeness and reliability have been measured as
function of point source flux density $x$. Completeness $C(x)$ is
defined as the ratio of the number of the matched sources
$N_{match}(x)$ to the number of the true sources $N_{true}(x)$.
Reliability $R(x)$ is defined as the ratio of the number of the
matched sources $N_{match}(x)$ to the number of the extracted sources
$N_{extract}(x)$:
\begin{eqnarray}\label{completentess_and_reliability}
C(x) = \frac{N_{match}(x)}{N_{true}(x)}; 
\nonumber \\
R(x) = \frac{N_{match}(x)}{N_{extract}(x)}; 
\end{eqnarray}

\begin{figure}
\figurenum{9}
\epsscale{1.0}
\plotone{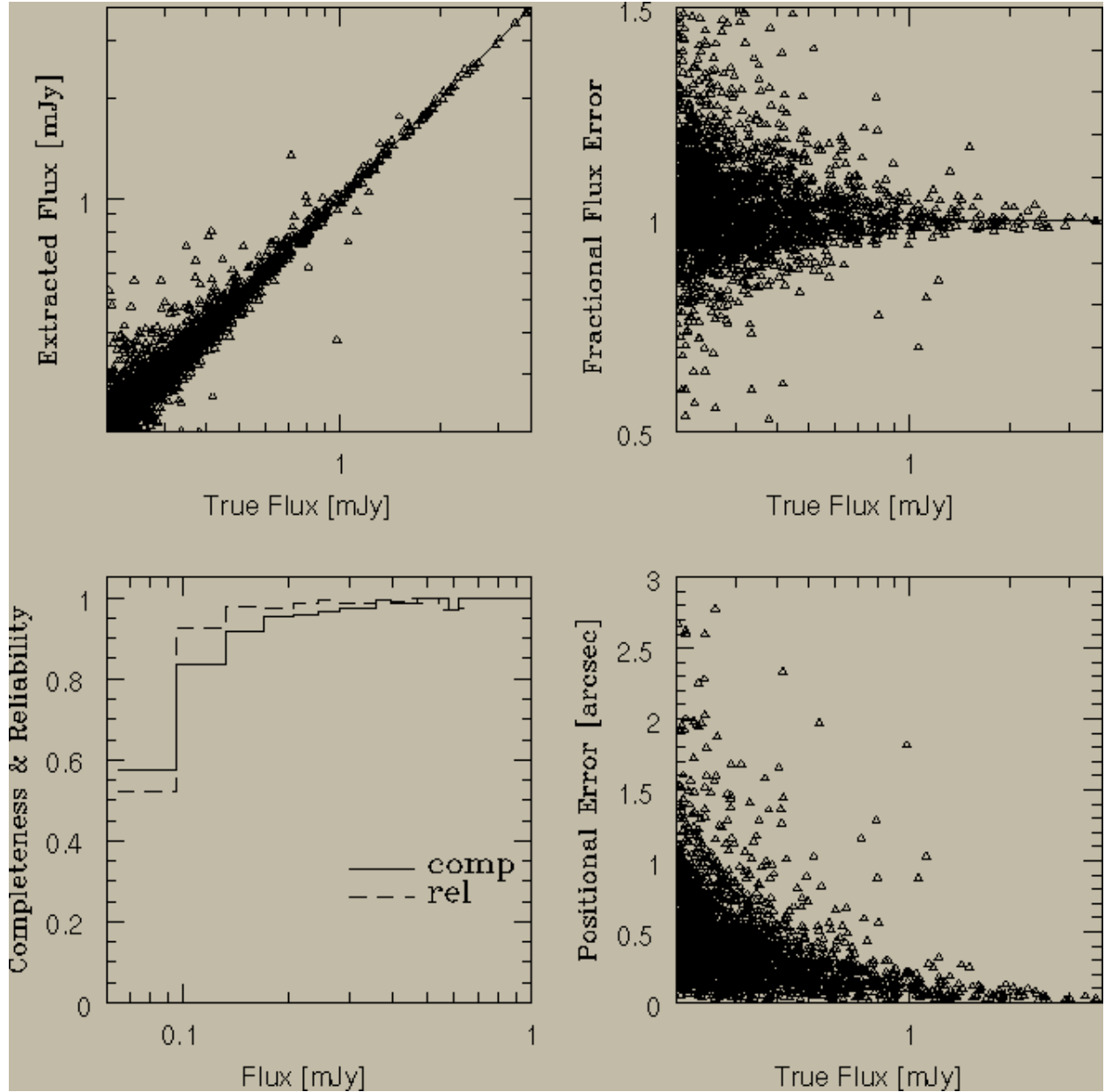}
\vspace{0mm}
\caption{Results of the simulations of the FLS main field for all 
extracted sources using MOPEX.
{\it Top Left:} Comparison of the profile fit fluxes from MOPEX with 
the true fluxes which are used in the simulated FLS main field 24 $\mu$m
images.  The straight {\it dotted line} represents the relation where
the MOPEX extracted flux equals the true flux.  {\it Top Right:} The 
fractional flux error, the ratio of the extracted 
flux to the true flux, is plotted as a function of true flux
 for the matched sources between the extracted and true point source
catalogues.  {\it Bottom Left:} Completeness and reliability measured
for the simulated images.  {\it Bottom Right:} The positional errors
as a function of the true flux. The positional errors are calculated
by comparing the true and measured positions in the simulated images.}
\label{simmain_apex}
\end{figure}

\begin{figure}
\figurenum{10}
\plotone{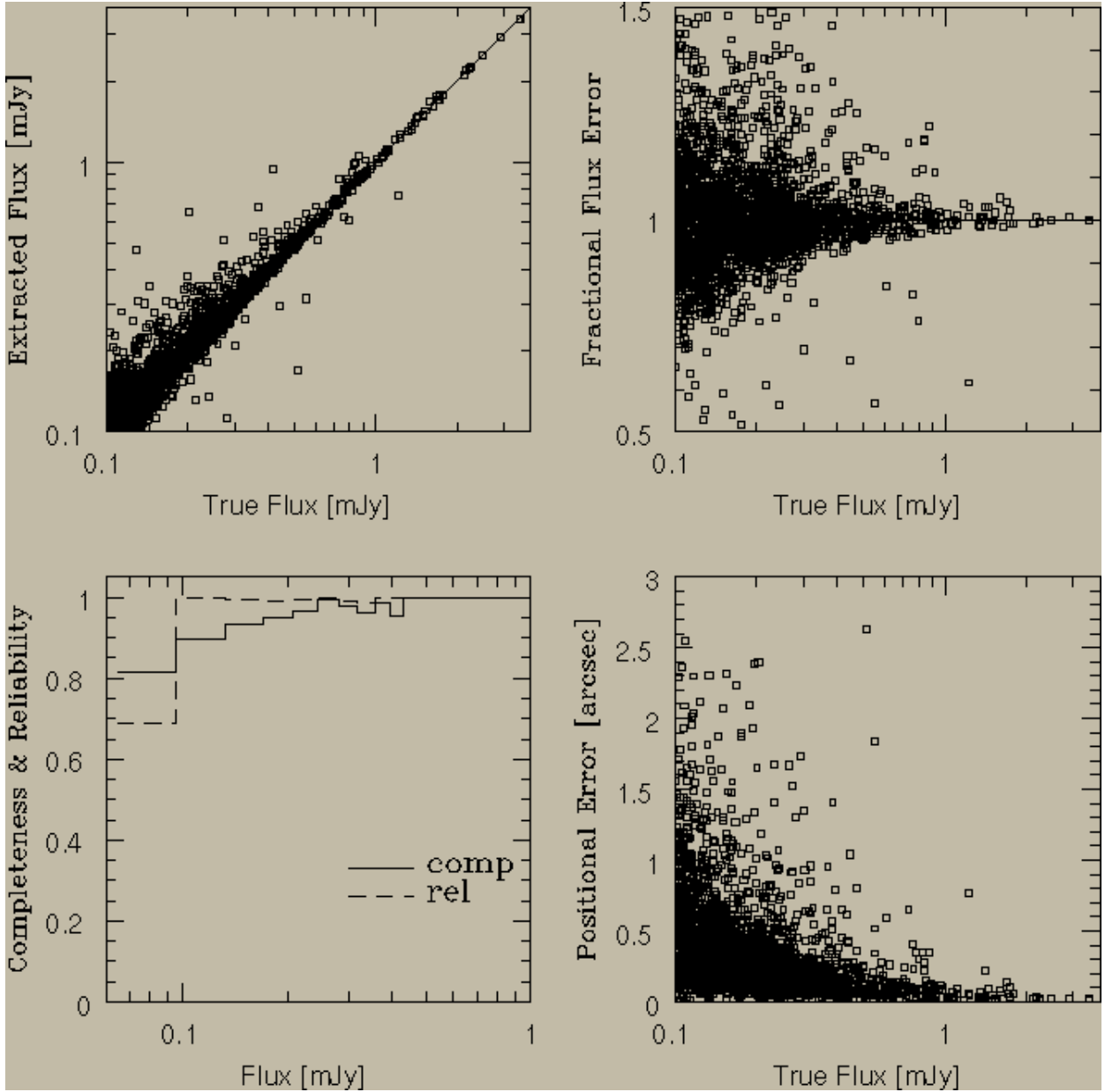}
\vspace{0mm}
\caption{Results of the simulations of the FLS verification strip for all 
extracted sources using MOPEX.
{\it Top Left:} Comparison of the profile fit fluxes from MOPEX with
the true fluxes which are used in the simulated FLS verification strip
24 $\mu$m images.  The straight {\it dotted line} represents the
relation where the MOPEX extracted flux equals the true flux.  {\it Top
Right:} The fractional flux error, the ratio of the extracted 
flux to the true flux, is plotted as a function of true flux
 for the matched sources between the extracted and true point
source catalogues.  {\it Bottom Left:} Completeness and reliability
measured for the simulated images.  {\it Bottom Right:} The positional
errors as a function of the true flux. The positional errors are
calculated by comparing the true and measured positions in the
simulated images.}
\label{simver_apex}
\end{figure}

The results of point source extraction of the simulation data are
shown in Figure~\ref{simmain_apex} (main survey) and
Figure~\ref{simver_apex} (verification strip). Comparison of the
profile fit fluxes from MOPEX with the true fluxes which are used
in the simulated FLS main field 24 $\mu$m images shows a tighter
relationship at faint fluxes for the deeper verification strip, as
expected.  Source extraction is 80\% complete at a flux of 0.11 (main)
and 0.08 mJy (verification). Within the 80\% completeness limit,
typical flux measurement errors are of the order of 10-15\% (depending
on SNR) or less and position accuracy is equal or better than 1
arcsec.

\begin{figure}
\figurenum{11}
\plotone{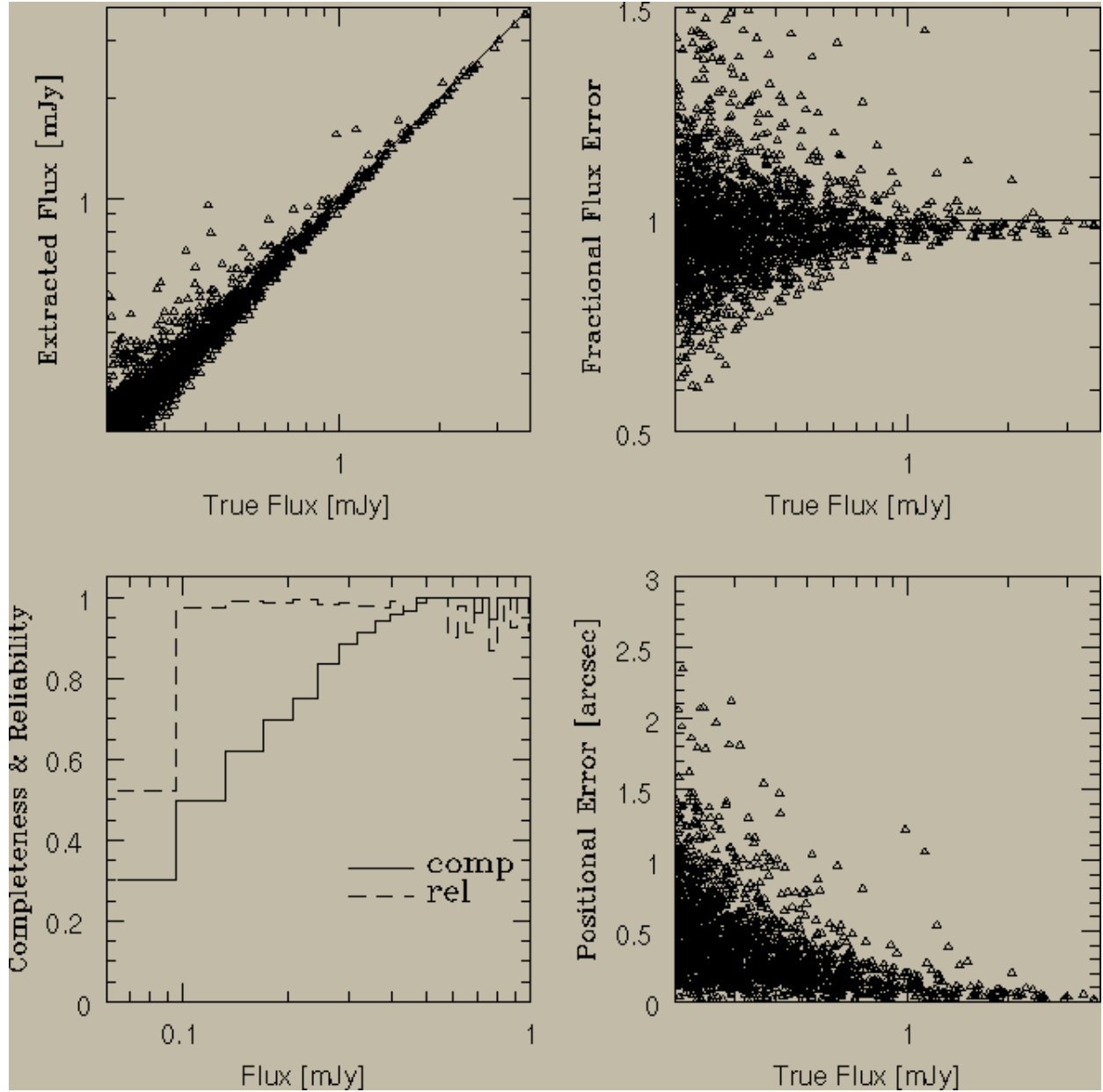}
\vspace{0mm}
\caption{Results of the simulations of the FLS main field for all 
extracted sources using DAOPHOT.}
\label{simmain_daophot}
\end{figure}

\begin{figure}
\figurenum{12}
\plotone{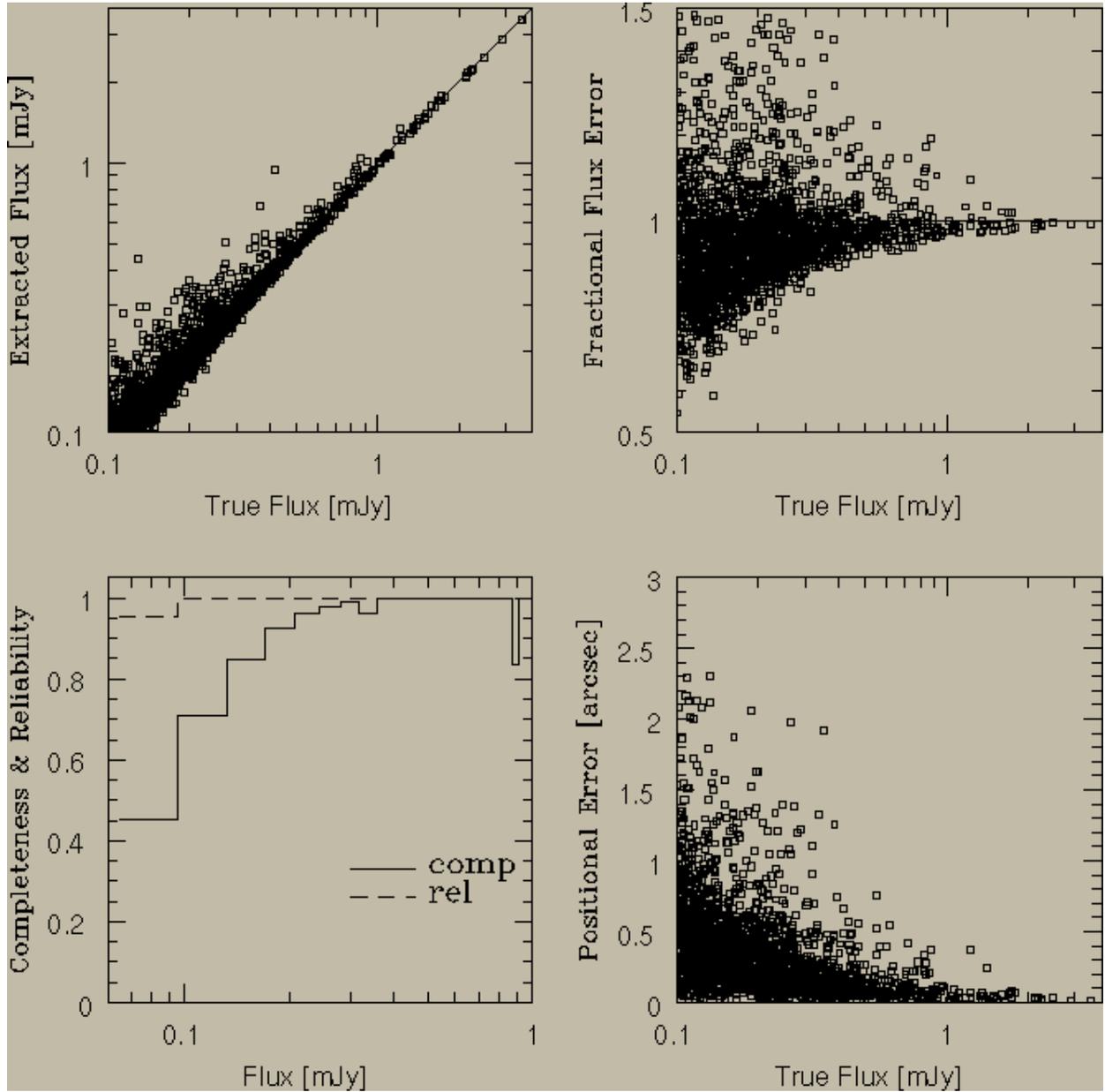}
\vspace{0mm}
\caption{Results of the simulations of the FLS verification strip for all 
extracted sources using DAOPHOT.}
\label{simver_daophot}
\end{figure}

We have performed a simple comparison with another source extraction
software, DAOPHOT. DAOPHOT was chosen because it is one of the best
known stellar photometry package \citep{DAOPHOT}. It was run as part
of IRAF\footnote{The Image Reduction and Analysis Facility (IRAF) is
distributed by the National Optical Astronomy Observatories, which are
operated by the Association of Universities for Research in Astronomy,
Inc., under cooperative agreement with the National Science
Foundation.} (version 2.12.1). The threshold (in sigma) for feature
detection was set to 1.5 and the fluxes were computed within an
aperture with radius of 13 pixels. The results of point source
extraction of the simulation data with DAOPHOT are shown in
Figure~\ref{simmain_daophot} (main survey) and
Figure~\ref{simver_daophot} (verification strip).

The results produced by MOPEX are arguably better than the ones
produced by DAOPHOT. 
The completeness and reliability of MOPEX extractions are
higher overall. That is, the reliability of DAOPHOT extraction is 
  slightly higher, but in our opition the gain
   in completeness in the much more significant than the loss in reliability.
Also MOPEX fluxes have no systematic offsets
at the higher end.

\subsection{Timing Results}\label{Benchmarking}
We have performed the timing test for MOPEX on a Solaris machine
(SunBlade 2500) with a CPU's clock rate of 1.5Ghz and 8GB of RAM.  We
ran MOPEX on the whole 4.4 square degree region and it took only 30
min to complete the point source extraction step (41559 sources extracted with SNR
$>$ 3). We also ran DAOPHOT on this image. The DAOPHOT counterpart of point
source extraction in MOPEX {\it nstar} took 43 min to complete.

\section{Conclusion}
We presented point source extraction with MOPEX, a package for
astronomical image processing developed at the {\it Spitzer} Science
Center. MOPEX performs point source extraction in two modes: mosaic
(single frame) and multiframe.  Point source extraction for well
sampled data and/or data with high depth of coverage should be done in
the mosaic mode.  We gave a description of the processing steps of
point source extraction in the mosaic mode and the main features that
contributed to accurate and efficient extraction.  Among them is the
non-linear matched filtering leading to improved detection of faint
point sources.  Passive and active deblending allow for successful
fitting of confused point sources.  Efficient methods of background
and noise estimation and the modified Simplex method contribute to the
computational speed of MOPEX.

MOPEX application was shown on the examples of MIPS24 and MIPS70 FLS
data and a Digital Sky Survey image. These included the low-level noise MIPS24 verification strip,
medium-level noise MIPS24 main field, and high-level noise MIPS70
verification strip.  In each case successful point source extraction
was evidenced by the quality of the residual images.  In order to
obtain quantitative evaluation of point source extraction we applied
MOPEX to simulated MIPS24 data.  We computed the completeness and
reliability of point source extraction, as well as the photometric and
astrometric accuracy.

For undersampled data with relatively low coverage point source
extraction is better done in the multiframe mode by simultaneously
fitting sources in the input frames instead of the mosaic image
created by coadding the input frames.  We will describe the multiframe
mode of point source extraction in the near future.
Another direction of exploration is applying
MOPEX to point source extraction in crowded fields.
This will require tuning the algorithm
and can potentially lead to some algorithm modification.

\acknowledgments We would like to express our gratitude to David
Frayer for making the MIPS70 mosaic and the PRF, and also to Dario
Fadda for making the MIPS24 mosaic.  This work is performed for {\it
Spitzer Space Telescope}, which is operated by the Jet Propulsion
Laboratory, California Institute of Technology, under a contact with
the National Aeronautics and Space Administration.
\clearpage

\appendix

\section{Non-linear Matched Filter}\label{app_matched}
Here we derive an expression for a non-linear filter based on the
notion of point source probability.  We assume that image ${\mathbf
s}$ observed by the detector consists of the point source contribution
${\mathbf x}$ convolved with the PRF and additive noise ${\mathbf n}$:
\begin{equation}\label{eq1}
{\mathbf s} = {\mathbf Hx} + {\mathbf n}.
\end{equation}
Here ${\mathbf H}$ is a translationally invariant matrix constructed
from the PRF: $ {\mathbf H_{ij}} = {\mathbf H_{i-j}} = PRF(i-j)$.  The
general problem is to estimate the probability of the point source
being at a particular pixel {\it i} given a measurement vector
${\mathbf s}$ in a certain window {\it W} surrounding this pixel.  The
size of the window {\it W} is determined by the size of the PRF.  We
assume that the point source and background noise are characterized by
the distribution functions $f_x({\mathbf s})$ and $f_n({\mathbf s})$,
correspondingly.  We consider two hypotheses for the pixel {\it i}.
The first hypothesis $h_1$ is that there is a point source at the
pixel, and the second $h_2$ (null hypothesis) is that there is not
a point source at the pixel.  The probability of the $k^{th}$
hypothesis conditioned on the measurement ${\mathbf s}$ is given by
the Bayesian theorem:
\begin{equation}\label{eq2}
P(h_k|{\mathbf s}) = \frac{f({\mathbf s}|h_k) P(h_k)}{f({\mathbf s})},
\end{equation}
where $P(h_k)$ is the a priori probability of the $k^{th}$ hypothesis,
$f({\mathbf s})$ is the probability density of observing the set of
pixel values ${\mathbf s}$. Assuming completeness of the hypothesis
set, i.e. $P(h_1) + P(h_2) = 1$, we obtain for $f({\mathbf s})$
\begin{equation}\label{eq3}
f({\mathbf s}) = f({\mathbf s}|h_1) P(h_1) + f({\mathbf s}|h_2) P(h_2).
\end{equation}
The probability density of measurement ${\mathbf s}$ under the null
hypothesis is given simply by the noise distribution function
$f_n({\mathbf s})$.  The probability density of measurement ${\mathbf
s}$ under the point source hypothesis is the result of integration
over all possible point source contributions ${\mathbf x}$ :
\begin{equation}\label{eq4} f({\mathbf s}|h_1) = \int d{\mathbf x}
f({\mathbf s}| {\mathbf x} ) f_x({\mathbf x}) = \int d{\mathbf x}
f_n({\mathbf s - Hx} ) f_x({\mathbf x}).
\end{equation}
Combining everything we obtain for the quantity in question:
\begin{equation}\label{eq5} P(h_1|{\mathbf s}) = (1 +
\frac{(1-P(h_1)) f_n({\mathbf s}) } {P(h_1) \int d{\mathbf x}
f_n({\mathbf s - Hx} ) f_x({\mathbf x}) })^{-1}
\end{equation}

In order to evaluate equation ~\ref{eq5} we assume that both the point
sources and the noise have zero-mean Gaussian distribution functions
with the variances $\sigma_x^2$ and $\sigma_n^2$ and are not
correlated spatially. We also assume that there is only one point
source in the window {\it W}. Under these assumptions the integration
in ~\ref{eq4} can be performed to yield:
 \begin{eqnarray}\label{eq6}
f({\mathbf s}|h_1) = \frac {1} { (\sqrt{2 \pi} \sigma_n)^W \sqrt{2\pi}
\sigma_x} \int dx \exp( -\frac{\sum_i(s(i) - PRF(i) x)^2} {2\sigma_n^2}
- \frac{x^2} {2\sigma^2_x} ) = \nonumber \\ \frac {\sigma_T} {(\sqrt{2
\pi} \sigma_n)^W \sqrt{2\pi} \sigma_x} \exp( -\frac {\sum_i s^2(i)} { 2 \sigma_n^2}
+ \frac { (\sum_i s(i) PRF(i) )^2} { 2 \sigma_n^4/\sigma_T^2} ).
\end{eqnarray}
Here 
\begin{equation}
\frac {1} {\sigma_T^2} = \frac {1} {\sigma_x^2} + \frac {\sum_i (PRF(i)
)^2} {\sigma_n^2}.
\end{equation}
After substituting equation ~\ref{eq6} in equation ~\ref{eq5} we
obtain the final expression for the point source probability:
\begin{equation}\label{eq7} P(h_1|{\mathbf s}) = (1 + \frac{(1-P(h_1))
\sigma_x } {P(h_1) \sigma_T} \exp(-\frac { \sum_i (s(i) PRF(i) )^2}
{2\sigma_n^4/\sigma_T^2} ) )^{-1}.
\end{equation}

\section{Modified Simplex Method}\label{app_simplex}
The original downhill simplex algorithm described in \citet{Nelder}
and \citet{O'Neill} minimizes a function of $N$ variables
by using the values of the
function at several vertices and trying to move away from the highest
vertex.  In our paper the function being minimized is the  goodness-of-fit measure
$\chi^2$ defined in equation~\ref{chi2}. There are four basic ways to move a vertex: reflection,
expansion, contraction and shrinkage.  

\begin{figure}
\figurenum{13}
\plotone{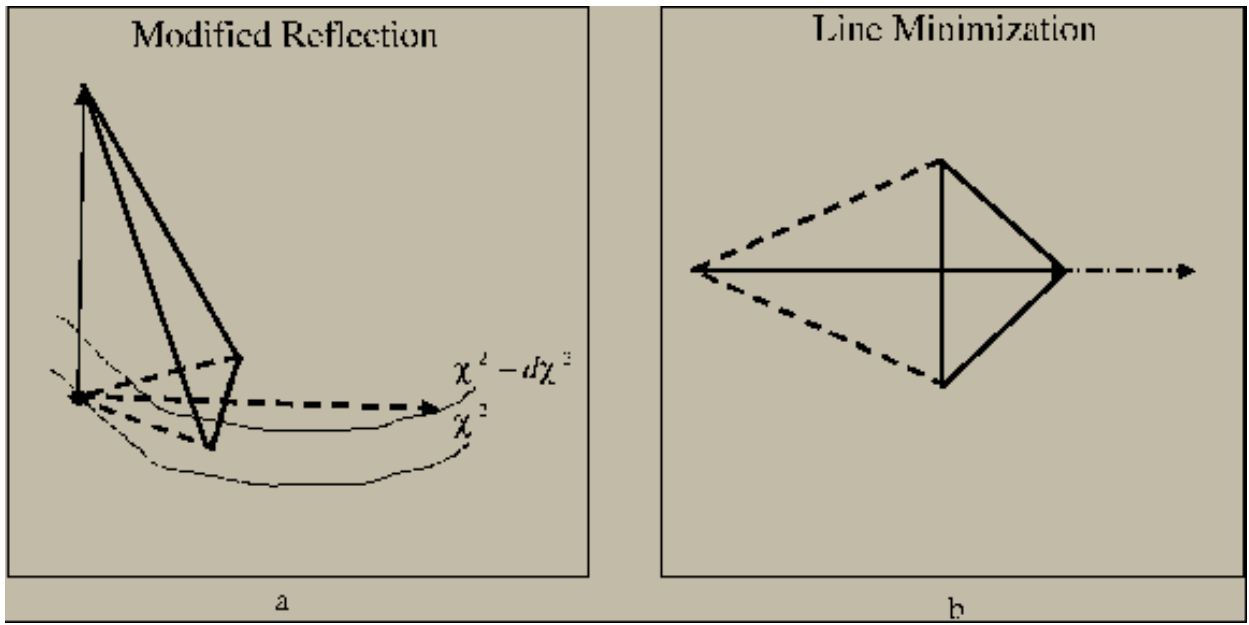}
\vspace{0mm}
\caption{Modifications of the Simplex method. a) The {\it dotted
lines} are lines of constant $\chi^2$.  The old simplex is shown with
{\it dashed lines}. The {\it dashed arrow} shows the reflection of the
highest vertex based on the old simplex. If the change $d\chi^2$ is
less than a user specified limit then the vertex is moved as indicated
by the {\it solid arrow} in a direction perpendicular to the original
reflection. b) The original simplex is shown with the {\it dashed
lines}. If reflection, shown here with the {\it dashed arrow}, results
in a point with a higher $\chi^2$, a point with the lowest $\chi^2$
is found along the line of reflection.}
\label{Simplex}
\end{figure}

We adopted the simplex algorithm with a
number of improvements.
The changes are illustrated in Figure~\ref{Simplex}. First, we modified
reflection.  If the change in $\chi^2$ is smaller than a
user-specified threshold, it is an indication that the reflection is
done almost parallel to the iso-$\chi^2$ lines. In this case an
attempt is made to replace the reflection with a move in a
perpendicular direction. The number of perpendicular directions
are equal to $2(N-1)$. The move is performed, if it results in a
$\chi^2$ lower than the one achieved by the reflection.

 Another modification is that contraction and
shrinkage have been replaced with the line minimization of $\chi^2$
along the unsuccessful reflection direction. I.e. if the reflection
results in point with higher than the original $\chi^2$, 
a point with the lowest $\chi^2$ is found on the line of the 
unsuccessful reflection.

Without the modifications the algorithm in its
original form very often was unable to find the global
minimum and remained stuck in of the local minima
and wandered away from the true point source location.

\clearpage 

\clearpage 

\clearpage 

\clearpage 

\clearpage 

\clearpage 

\clearpage 

\clearpage

\clearpage

\clearpage

\clearpage

\clearpage 

\clearpage 

\clearpage

\begin{deluxetable}{crrrrrrrrrr}
\footnotesize
\tablecaption{Test of passive deblending.
The average execution time $t_{ex}$, 
positional $\delta_R$ and flux $\delta_f$ 
errors are shown as a function
of the number of sources in a cluster $N_p$.
\label{MIPS_Sim_P}}
\tablewidth{0pt}
\tablehead{
\colhead{$N_p$} &
\colhead{ 2}&
\colhead{ 4} &
\colhead{ 6} &
\colhead{ 8} &
\colhead{ 10} & 
\colhead{ 12}&
\colhead{ 14} &
\colhead{ 16} &
\colhead{ 18} &
\colhead{ 20}  
}

\startdata
$t_{ex}$(sec)              &0.07 &0.15 &0.43 &0.90 &1.8 &3.3 &5.5 &9.3 &15  &20  \\ \\
$\delta_R$(10$^{-2}$pixels)  &0.3  &0.5  &0.8  &1.7  &2.4 &2.6 &2.9 &3.5 &4.3 &5.1  \\ \\
$\delta_f$(10$^{-2}$)        &0.5  &0.8  &1.1  &1.6  &1.9 &2.2 &2.4 &2.7 &2.9 &3.3   \\ 
\enddata

\end{deluxetable}

\end{document}